\documentclass[conference]{IEEEtran}
\usepackage[pdftex]{graphicx}
\usepackage[cmex10]{amsmath}
\newtheorem{thm}{Theorem}[section]

\newtheorem{lem}[thm]{Lemma}
\newtheorem{defn}[thm]{Definition}

\begin{document}
%
\title{Stable, scalable, decentralized P2P file sharing with non-altruistic peers}

\author{\IEEEauthorblockN{Barlas O\u{g}uz, Venkat Anantharam}
\IEEEauthorblockA{Department of Electrical Engineering and Computer Sciences\\
University of California Berkeley\\
Email: \{barlas, ananth\}@eecs.berkeley.edu}
\IEEEauthorblockN{Ilkka Norros}
\IEEEauthorblockA{VTT Technical Research Centre of Finland
\\
Email: ilkka.norros@vtt.fi }}

\maketitle

\begin{abstract}
P2P systems provide a scalable solution for distributing large files in a network.  The file is split into many chunks,
and peers contact other peers to collect missing chunks to eventually complete the entire file.  The so-called 
`rare chunk' phenomenon, where a single chunk becomes rare and prevents peers from completing the file, is a threat to the
stability of such systems.  Practical systems such as BitTorrent overcome this issue by requiring a global search for the
rare chunk, which necessitates a centralized mechanism.  We demonstrate a new system based on an 
approximate rare-chunk rule, allowing for completely distributed file sharing while retaining scalability and stability.
We assume non-altruistic peers and the seed is required to make only a minimal contribution.
\end{abstract}
\IEEEpeerreviewmaketitle

\section{Introduction}
The marvel of peer-to-peer (P2P) networks is their scalability and robustness.  Both of these attributes are due in turn to
the distributed nature of such systems.  In a P2P file sharing system such as \emph{BitTorrent} \cite{cohen2003incentives}, a large file is split into many
chunks.  A peer who downloads a chunk can immediately start uploading that chunk to other peers, contributing to the
resource pool of the sharing network.  

To ensure the availability of all chunks of the file at all times, at least 1 seed (who has the complete file) is assumed
to stay in the network at all times.  However, in an open system, where peers are arriving according to a random arrival
process, the presence of a single seed does not guarantee stability.  It has been observed and demonstrated through various
analytical models (\cite{norros2006flash,norros2009stability,norros2007uncoordinated,mathieu2006missing})
 that if the peers contact each other and download chunks in a purely random fashion, a single chunk might be driven to near 
extinction, causing peers to stay in the system for a long time and driving the number of peers in the system to infinity.
In this scenario, the rare chunk is difficult to obtain, because there is a high probability that the randomly contacted peer 
does not have it, and peers who have the rare chunk tend to leave the system quickly since they probably already have every 
other chunk.  Thus the rare chunk becomes progressively rarer as new peers accumulate.

It is known that the rare chunk issue can be avoided through altruistic behavior of the peers \cite{mathieu2006missing,qiu2004modeling}.  
In this model, peers who complete the file stay in the system for an additional random time to aid the remaining peers.
It has recently been proven \cite{zhu2011stability} that it is sufficient for peers to remain in the network for a
time that is on average equal to the time it takes to download a single chunk.  Unfortunately in real networks, 
the altruistic peer assumption may not hold.  

A similar stabilizing effect is observed if the upload capacity of the seed is large enough to maintain a balanced
 chunk distribution in the system \cite{massoulie2005coupon,zhu2011stability}.  However, in such scenarios, 
the demand on the seed scales with the number of peers in the system, reducing the scalability of the protocol.
 
\emph{BitTorrent} addresses the `rare chunk' issue by forcing peers to download the rarest chunk first.  This rule necessitates
a centralized search in the network to track the rarest chunk, and the peers who possess that chunk.  The tracker is the
only centralized piece in the BitTorrent protocol, and presents a single point of failure for the system.  It is desirable
to replace this mechanism by a distributed rule that approximates the `rarest chunk first' rule for downloads.

In this paper, we present a new P2P file sharing protocol that is provably stable.  Our protocol is completely distributed, and 
fully scalable, avoiding the pitfalls of a centralized tracker or a privileged seed.  Moreover, we assume that peers who complete
the download leave immediately. Stability depends on a probabilistic local rule that peers follow to approximate prioritizing the rare chunk.  

We model an open system, where peers are arriving according to a Poisson process.  In our model, the current chunk
profiles of all the peers in the system defines the state for a Markov process.  This is in line with the model described
in \cite{massoulie2005coupon}. In contrast with deterministic fluid models such as the one used in \cite{qiu2004modeling},
we attempt to directly prove the stability of the dynamic system.  This approach is more reassuring, as a well defined
relationship between the stability of the dynamical system and that of the fluid models is yet to be formulated \cite{norros2009stability}.
Our model and assumptions are detailed in section \ref{model}. 

The protocol is presented in section \ref{solution}.  It is a modification of the `majority' rule that was first 
proposed in \cite{reittu2009stable,norros2008urn}, where
the authors used simulations and large system limits to argue that the rule leads to a stable system.  We have since
been able to formally prove stabililty for the special 2-chunk case \cite{norros2009stability} (see section \ref{performance} for a brief discussion),
however stability for the general case remains a conjecture.  Our proposal implements a stricter rule to keep
the rare chunks in the system longer, and allows us to prove stability in general for any number of chunks and any
arrival rate.  The proof, presented in section \ref{proof}, employs an unconventional Lyapunov function, which is
the main contribution of this paper together with the new protocol.
The form of the Lyapunov function is quite novel and might be useful in proving the stability of similar algorithms.

\section{Problem definition}\label{model}
A file is divided into $k$ chunks, to be distributed in a P2P system.  Peers enter the system at Poisson rate $\lambda$
and leave immediately upon receiving all $k$ chunks.  We make the following assumptions:
\begin{itemize}
 \item There is always exactly $1$ seed in the system (denoted by $s$), who has all the chunks.  
 \item At a Poisson rate of 1, each peer can sample randomly with replacement up to 3 peers from the current population $\cup \{s\}$.
(A peer is allowed to sample itself.)
 \item At the time of sampling, the peer can choose to download at most $1$ chunk which it does not already have, 
   but shows up in the sample.  The download is assumed to happen instantaneously.
 \item A newly arriving peer arrives with no chunks.
\end{itemize}
Let $S$ be the total number of peers in the system (including the seed). Note that each peer 
(including the seed) can be sampled, on average, at most by three other peers per unit time.
Therefore the average upload bandwidth per peer is bounded, and does not scale with $\lambda$ or $S$.

We seek a rule by which a peer can decide which chunk (if any) to download at each time slot from the current sample.
This rule is assumed to be a function of the peer's current chunk profile, and the profiles of the sampled peers.  
(The rule could possibly depend on the past observations of the peer, as well as $\lambda$ and $k$.  However the rule
we propose will not depend on these.)  We require that the chosen rule stabilizes the stochastic system in the Lyapunov sense.

\section{Solution: Common chunk protocol}\label{solution}
Let $S_i = \text{\# peers who have chunk $i$ including the seed}$, $i\in \{1, \ldots, k\}$. 

$S_0 =  \text{\# peers who have no chunks}$.

$\bar S_i = S- S_i$.


$\bar T_i = \text{\# peers who have only chunk $i$ missing}$.  

\begin{defn}
 A chunk in a sample of $3$ peers is \emph{rare} if exactly $1$ peer in the sample has that chunk.
 A chunk is called a \emph{match} if it is contained in the sample but not in the sampling peer's profile.
\end{defn}

We define the rule as follows:
\begin{itemize}
 \item Peers with no chunks sample 3 peers at random and choose to download a chunk that is a rare match.
  If there is more than $1$, they pick randomly among them.  If there are no rare matches, the peer skips this time slot
  without downloading.
 \item Peers who have more than zero, but less than $k-1$ chunks sample only 1 peer at random, and download a chunk at random
  among those that match (no rare match required). Skip if there is no match.
 \item Peers who have $k-1$ chunks sample 3 peers at random.  Download only if every chunk that the peer has appears at least
twice in the sample, and there is a match.  Otherwise skip without downloading.
\end{itemize}

Roughly, the first item is meant to stop arriving peers from acquiring a common chunk as their first chunk.  The last item
attempts to keep rare chunks from leaving the system.  This should balance out the chunk distribution in the system and 
provide stability.  

Note that by sampling only 3 peers, we are requiring the bare minimum that allows a majority rule.
By sampling more peers, one could clearly do better, however our main purpose here is to demonstrate that stability is possible
even in this restricted setting.  We discuss sampling more peers and other performance enhancing heuristics in section 
\ref{performance}.

In this paper, we model the proposed system by a Markov process with state space $\mathcal{X}$ described by the peers currently
in the system and their chunk profiles.  The description of the state space is essentially identical to that in 
\cite{massoulie2005coupon}.  Let $q(\mathbf{x},\mathbf{x'})$ denote the entries of the generator matrix of this Markov
process.  For any function on the state space,
\begin{defn}
 The \emph{drift} $\Delta f(\mathbf{x})$ of a function $f(\mathbf{x})$ is defined as
\[ \Delta f(\mathbf{x}) = \sum_{\mathbf{x'} \ne \mathbf{x}} q(\mathbf{x},\mathbf{x'})(f(\mathbf{x'})-f(\mathbf{x})). \]
\end{defn}

We use the following well known tool for the proof:
\begin{thm}\label{lyapunov}[Foster-Lyapunov]
 Let $L$ be a function on the state space with drift $\Delta L$.  Let $L\ge 0$ and let
$\{L \le l \}$ be a finite set for any finite constant $l > 0$.  If for an $\epsilon > 0$, 
$\Delta L < -\epsilon$ on the set $S > c$, for a suitably chosen constant $c$, then the Markov process is positive recurrent.
\end{thm}

\section{Proof of stability}\label{proof}
 We will split the proof into two cases according to whether $\lambda \le \frac{1}{3k}$ or $\lambda > \frac{1}{3k}$.
In each case, we will show the stability of this system by demonstrating a Lyapunov function for it. 

Let $r$ be the total rate of downloads. $dS_i^+$ is the virtual rate (stochastic intensity) at which a peer with no chunks downloads chunk $i$, 
and $d\bar S_i^-$ is the virtual rate at which a peer who is lacking only chunk $i$ downloads $i$ and leaves the system.
We will need the following lemmas:

\begin{lem}\label{lemma1}
 $r \ge \frac{S r_0^2}{6k^2}$ where $r_0 = \sum_i dS_i^+$.
\end{lem}
\begin{proof}
 Write $\bar S_i = S_0 + \bar B_i + \bar T_i$ to define $\bar B_i$. $\bar B_i$ is the number of peers who lack 
chunk $i$ and have at least 1 and at most $k-2$ chunks. We can write 
\[ r \ge S_0 d(S_0) + \bar T_i d(\bar T_i) + \bar B_i d(\bar B_i) \]
where $d(.)$ denotes the virtual rate of downloads for an individual peer in each group.  By definition, 
$d(S_0) = \sum_i dS_i^+ \ge dS_j^+$ for any $j$.  Also
\[d(\bar T_i) \ge \frac{3\bar T_i ^2 S_i}{ S^3} = \frac{\bar T_i ^2}{ \bar S_i^2}\frac{3 \bar S_i ^2 S_i}{ S^3} \ge \frac{\bar T_i ^2}{ \bar S_i^2}dS_i^+\]
where the last inequality follows because $\frac{3 \bar S_i ^2 S_i}{ S^3}$ is the probability of a rare match, but $dS_i^+$
might be smaller due to the possibility of multiple rare matches.  It is left to note $d(\bar B_i) \ge \frac{dS_i^+}{3}$.  To
argue this, note that $d(\bar B_i)$ is the total virtual rate of downloads per peer for peers in $\bar B_i$, and a sample of chunk $i$ is sufficient
to result in a download (even if chunk $i$ is not chosen to be downloaded).  The probability of sampling chunk $i$ in 1 go
is at least $\frac{1}{3}$ the probability of sampling chunk $i$ in 3 tries, which in turn at least as large as $dS_i^+$. We can 
now write
\[ r\ge (S_0 + \frac{1}{3}\bar B_i)dS_i^+ + \frac{\bar T_i^3}{\bar S_i^2} dS_i^+ \ge \frac{\bar S_i}{6} dS_i^+.\]
Here we argue that $x+y+z = 1 \implies x + \frac{y}{3} + z^3 \ge \frac{1}{6}$.
 \[ \frac{\bar S_i}{S} \ge dS_i^+. \]
Therefore 
\[ r \ge \max_i \frac{S (dS_i^+)^2}{6} \ge \frac{r_0^2 S}{6k^2}.\]
\end{proof}

\begin{lem}\label{lemma2}
 \[r \ge \min_i \frac{S_i}{2k^3}, \quad \text{if } S\ge 12 .\] 
\end{lem}
\begin{proof}
Argue as in the previous lemma that for $S_0$, a rare match of $i$ is sufficient for a download. For $\bar B_i$, a simple
match is sufficient and $\frac{S_i}{S} \ge \frac{\bar S_i^2 S_i}{S^3}$.  Therefore
 \[ r\ge (3S_0 + \bar B_i)\frac{\bar S_i^2 S_i}{S^3} + 3\frac{\bar T_i^3}{\bar S_i^2}\frac{\bar S_i^2 S_i}{S^3} \ge \frac{2\bar S_i^3 S_i}{3S^3}, \text{  for any $i$}.\]
Here we argue that $x+y+z = 1 \implies x + \frac{y}{3} + z^3 \ge \frac{2}{9}$. Since $\max_i \bar S_i \ge \frac{S-1}{k}$, we have the result when $S\ge 12$.
\end{proof}

First consider the case $\lambda \le \frac{1}{3k}$.

We propose:
 \[ L_1 = \sum_i \bar S_i \]
The drift of $L$ is
\[ \Delta L_1 = k\lambda - r \le \frac{1}{3} - r. \]
From lemma \ref{lemma2}, we know 
\[ r > \frac{\bar S_i^3 S_i}{2S^3}, \text{  for any $i$}.\]
Picking $i$ to be the rarest chunk, we observe that $r > \frac{1}{3}+\epsilon$ whenever $S > 3k^3$.

Now assume $\lambda \ge \frac{1}{3k}$.  

 We propose:
 \[ L = C\underbrace{\sum_i \bar S_i}_{L_1} +  \underbrace{\sum_i \frac{S}{e^{S_i}} + \frac{S}{e^{S_0}}}_{L_2} \]
where $C$ is a constant to be chosen later.

We will calculate the drift of $L$ in two parts.
A download of chunk $i$ decreases $\bar S_i$ by one and leaves all other $\bar S_j$ unchanged. A new arrival 
increases each $\bar S_i$ by one. Therefore
\[ \Delta L_1 = k\lambda - r. \] 
Also 
\begin{align*}
 \Delta L_2 &\le \sum_{i} \frac{\lambda}{e^{S_i}} -\lambda\left(\frac{S}{e^{S_0}} -\frac{S+1}{e^{S_0+1}}\right) \\
 &- \sum_i S_0 S dS_i^+ \left[ \frac{1 }{e^{S_i}} - \frac{1}{e^{S_i+1}} + \frac{1 }{e^{S_0}} - \frac{1}{e^{S_0-1}}\right] \\
&+ \sum_i \bar T_i d\bar S_i^- \sum_{j\ne i} \left(\frac{S-1}{e^{S_j-1}} - \frac{S}{e^{S_j}}\right)
\end{align*} 

The first two terms correspond to the the arrival of a new peer.  The second term is the drift due to a peer with no chunks
downloading chunk $i$.  The last term corresponds to the event where a peer leaves the system after having downloaded chunk $i$.
 
The inequality is due to the fact that we omitted
terms corresponding to transitions which keep $S$ and $S_0$ constant, and in the last set of terms, for each individual 
$i$ we ignored the terms corresponding to $S_i$ and $S_0$.  These transactions can only decrease $L_2$.

Since $d\bar S_i^- \le \min_{j \ne i} \frac{3 S_j^2}{S^2}$ and $\bar T_i \le S_j, \quad \forall j\ne i$, the last term satisfies
\begin{align}
\sum_j &\left(\frac{S-1}{e^{S_j-1}} - \frac{S}{e^{S_j}}\right) \sum_{i\ne j} \bar T_i d\bar S_i^- \notag \\ 
&\le \sum_j 3(e-1)(k-1)\frac{\bar S_j^3}{S^2}  \frac{S}{e^{S_j}} < \frac{8k^2}{S} . \label{top}
\end{align}
We get 
\begin{align*}
\Delta L_2 &< \sum_{i} \frac{\lambda}{e^{S_i}} + \frac{\lambda}{e^{S_0}} -\frac{\lambda (e-1) S}{e\cdot e^{S_0}}  \\
&+ \sum_i (e-1)S_0 S dS_i^+\left[\frac{1}{e^{S_0}}-\frac{1}{e\cdot e^{S_i}} \right]  + \frac{8k^2}{S} 
\end{align*}
Since $\sum_{i} \frac{\lambda}{e^{S_i}} + \frac{\lambda}{e^{S_0}} < (k+1)\lambda$, we may write
\begin{align*}
\Delta L &< \frac{8k^2}{S}+\lambda+(C+1)k\lambda - Cr - \frac{\lambda (e-1) S}{e\cdot e^{S_0}} \\
 &+ \sum_i (e-1)S_0 S dS_i^+\left[\frac{1}{e^{S_0}}-\frac{1}{e\cdot e^{S_i}} \right] 
\end{align*}

Now we are ready to show
\begin{thm}\label{main}
 \[ \Delta L < -\epsilon  \]
with $\epsilon > 0$ and $C = 108ek^3$ whenever $S > 4Ck e^{3C\lambda k^2 e^{6\lambda k^4}}$.
\end{thm}
\begin{proof}
Since we assume that $\frac{8k^2}{S} < 1$, we are left with 5 terms which can be written as
\begin{align*}
\Delta L &< [1+\lambda+(C+1)k\lambda] - Cr - \frac{\lambda (e-1) S}{e\cdot e^{S_0}} \\
 &+ \frac{(e-1)S_0 r_0 S }{e^{S_0}} - \sum_i \frac{(e-1)S_0 S dS_i^+ }{e\cdot e^{S_i}} 
\end{align*}
\begin{itemize}
 \item If $r \ge 2k\lambda$:
 \begin{itemize}
  \item If $\frac{(e-1)S_0 S r_0}{e^{S_0}} \ge \frac{Cr}{3}$:
    Then $r_0 \le \frac{S_0}{6eke^{S_0}}$ by lemma \ref{lemma1} and $r_0 S_0 \le \frac{S_0^2}{6eke^{S_0}} < \frac{1}{6ek}$.
   The third and fourth terms give at most $\frac{-(e-1)(\lambda-1/6k)S}{e\cdot e^{S_0}}$, which is negative.  Since $r \ge 2k\lambda$, we're done.
  \item If $\frac{(e-1)S_0 S r_0}{e^{S_0}} < \frac{Cr}{3}$, we have
 \[ \Delta L < 1+\lambda+(C+1)k\lambda - \frac{4}{3}Ck\lambda < -\epsilon .\]
 \end{itemize}
 \item If $r < 2k\lambda$, then by lemma \ref{lemma2}, $\exists S_{i^*} < 6k^4 \lambda$.
      Since $dS_i^+ > \frac{3\bar S_i^2 S_i}{kS^3}$, the last term is at most $-\frac{ S_0 }{k e^{6k^4 \lambda}}$.
Here we used the bound $\frac{3(e-1)\bar S_{i^*}^2}{eS^2} > 1$.
 \begin{itemize}
  \item If $\frac{(e-1)S_0 S r_0}{e^{S_0}} \ge \frac{Cr}{3}$:
    Then $r_0 S_0 \le \frac{S_0^2}{6eke^{S_0}} < \frac{1}{6ek}$. The third and fourth terms give at most $\frac{-(e-1)(\lambda-1/6k)S}{e\cdot e^{S_0}} \le \frac{-(e-1)\lambda S}{2e\cdot e^{S_0}}$, which is negative.
    If $S_0 \ge 3C\lambda k^2 e^{6k^4 \lambda}$, the last term is less than $-\frac{3}{2}Ck\lambda$, and
    \[ \Delta L < 1+\lambda+(C+1)k\lambda - \frac{3}{2}Ck\lambda < -\epsilon .\]
    Else $S_0 < 3C\lambda k^2 e^{6k^4 \lambda}$, so we would have $-\frac{(e-1)\lambda S}{2e\cdot e^{S_0}} < -\frac{2(e-1)}{e}Ck\lambda$ and 
  \[ \Delta L < 1+\lambda+(C+1)k\lambda -\frac{2(e-1)}{e}Ck\lambda < -\epsilon \]
since $\frac{2(e-1)}{e} > \frac{5}{4}$.
\vspace{0.3cm}
  \item If $\frac{(e-1)S_0 S r_0}{e^{S_0}} < \frac{Cr}{3}$, we can omit the second and fourth terms which add up to
  $\frac{(e-1)S_0 S r_0}{e^{S_0}}-Cr < 0$.  Again by the same
  reasoning as above, if $S_0 \ge 3C\lambda ^2 e^{6k^4 \lambda}$, the last term is less than $-\frac{3}{2}Ck\lambda$, and
    \[ \Delta L < 1+\lambda+(C+1)k\lambda - \frac{3}{2}Ck\lambda < -\epsilon .\]
     Else $S_0 < 3C\lambda k^2 e^{6k^4 \lambda}$, so we would have $-\frac{(e-1)\lambda S}{e\cdot e^{S_0}} < -\frac{2(e-1)}{e}Ck\lambda$ and 
  \[ \Delta L < 1+\lambda+(C+1)k\lambda -\frac{2(e-1)}{e}Ck\lambda < -\epsilon \]
since $\frac{2(e-1)}{e} > \frac{5}{4}$.
 \end{itemize}
\end{itemize}

\end{proof}

It is clear that both Lyapunov functions that are used satisfy the properties of theorem \ref{lyapunov}.  We conclude
that the proposed system is positive recurrent for any value of $\lambda$.

\section{Performance}\label{performance} 
From a performance point of view, some aspects of the protocol may strike the reader as inefficient.  In particular,
the rule for leaving the system is quite strict, and may cause substantial delay for the peers that have all but one chunk.
Consider a state where most of the peers have a few or no chunks.  A sample of $3$ peers needs to contain at least $2k-1$
chunks (1 for the missing chunk, 2 each for the rest) for a peer to be able to leave the system.  Therefore a peer with 
$k-1$ chunks will need to wait in the system, until the system becomes more saturated.


On the other hand, this ensures availability of all chunks to other peers, and reduces starvation in the network.  This
rule can be interpreted as forcing a degree of altruistic behavior and has a similar effect in terms of stability.  

\subsection{$m$-sampling}
The rule for the peers that have all but one chunk can be eased as follows.  A peer in $\bar T_i$ samples $m$ peers at 
random with replacement instead of $3$.  Allow a download only if each chunk other than $i$ is observed at least twice
in the sample of $m$ peers.  This would ensure none of the chunks which leave are rare.  Sampling more peers increases
the complexity of the system (decreases locality), but allows for a more efficient search (peers could leave earlier).  
One could pick $m$ to strike a good trade-off between complexity and performance.

The proof of stability generalizes to this case with little modification. In (\ref{top}), note that 
$d\bar S_i^- \le \min_{j \ne i} {m \choose{2}} \frac{S_j^2}{S^2}$.  Therefore the last term would be bounded by 
${m \choose{2}} \frac{8k^2}{3S}$, which can be bounded by $1$, provided we modify the bound on $S$ in theorem \ref{main}
with ${m \choose{2}}$.  The rest of the proof goes through unaltered.

\subsection{Rare chunk rule}
The original rule proposed in \cite{reittu2009stable,norros2008urn} was as follows:
All peers sample 3 other peers with replacement, and download only if there is a rare match.  As noted before, this
is a minimalist approach to approximation by a majority rule. The difficulty that arises in trying to prove the stability
of this system is that the majority rule does not in general favor the rare chunk, but rather inhibits the common chunk.  
These two goals turn out to be identical in the special 2-chunk case, for which it has been possible to find a Lyapunov function:

\begin{thm}
 \[ L = 2(2S_0 + S_1 + S_2) + (S_1-S_2)^2, \quad S > 30\lambda (20\lambda +1)^2 \]
is a valid Lyapunov function for the 2-chunk system with the rare chunk rule described above.
\end{thm}

In the interest of keeping our focus, we omit the proof of this result.  We will only remark that the first term
will be decreasing whenever there is sufficient balance in the system, and the second term turns out to be always
decreasing due to the rare chunk rule, and makes up for the increase in the first term when the system is in severe imbalance.

\subsection{Simulations}
We compare our proposed algorithm with the parameter $m$ taking the values $\{3,5,10\}$, $m=3$ being the original protocol
proposed in section \ref{solution}, with the rare chunk rule.

\begin{figure}
 \includegraphics[width=0.5\textwidth]{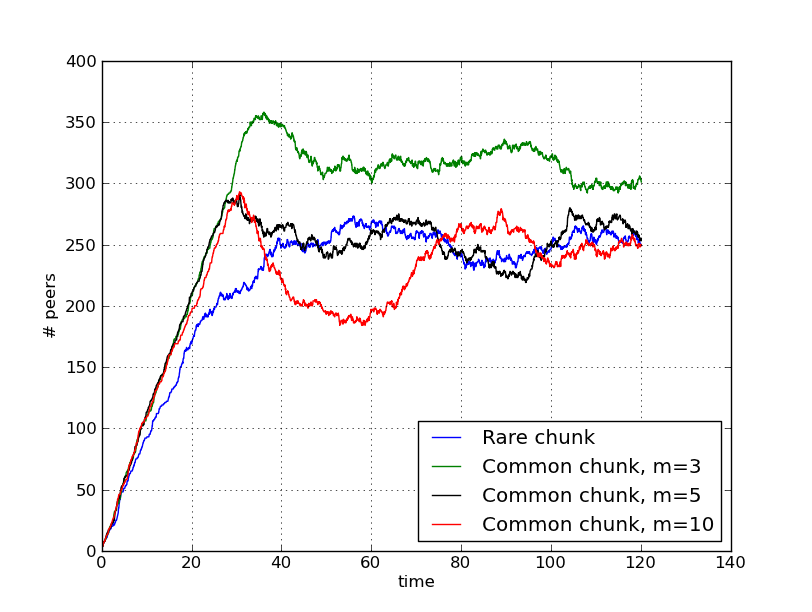}
 \caption{Reaching steady state from an empty system. $k=20$. $\lambda=10$.}
 \label{fig1}
\end{figure}

\begin{figure}
 \includegraphics[width=0.5\textwidth]{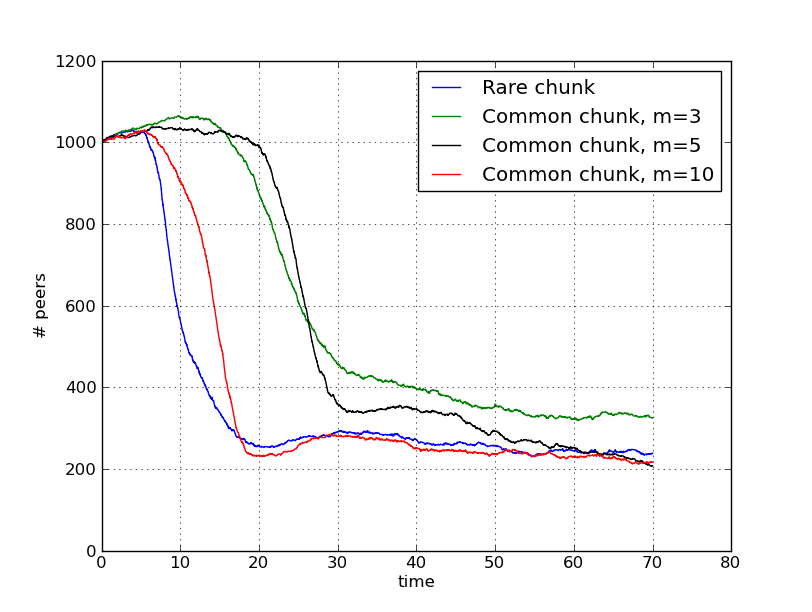}
 \caption{Relaxing from severe imbalance. 1000 peers all lack a single chunk at initialization. $k=20$. $\lambda=10$.}
 \label{fig2}
\end{figure}

Figure \ref{fig1} shows a system with 20 chunks and $\lambda = 10$.  At time 0, only the seed is present.  We can see all
four systems reaching a stable state.  The total number of peers for $m=5$ and $m=10$ behave roughly similar to the
simple rare chunk algorithm, where $m=3$ hovers slightly above the others due to the stricter rule keeping peers in the system
for a longer time.  The same behaviour is observed in figure \ref{fig2}, where all systems relax in a similar manner from an
initial population of 1000 peers, all of which lack the same chunk.

As reported in \cite{norros2009stability,reittu2009stable}, the rare chunk rule seems to provide stability despite the lack of a conclusive proof in this
direction.  On the other hand, the newly proposed protocol performs competitively (and even more so with a suitably chosen
parameter $m$) while having the advantage of a formal stability guarantee.

One should note that the total queue size is not necessarily the best indicator of performance.  It would be interesting
to see whether there are differences in the mean sojourn time of the peers with different protocols, and how such
metrics might be affected by the parameter $m$.  These topics will be subjects of further research.

\section{Conclusions}
Peer-to-peer schemes such as \emph{BitTorrent} have been remarkably successful in revolutionizing the way files are
spread in a network.  Still, it is desirable to completely decentralize such protocols to avoid the pitfall of a 
single tracker.  Naive attempts at such schemes have been plagued with the `rare chunk' syndrome, which causes instability.
While it has recently been shown that relatively minor altruistic behavior or a powerful seed can stabilize such systems,
these properties are usually a luxury in real world networks.  

In this paper we have demonstrated that a completely decentralized, stable peer-to-peer network is possible, even with completely
non-altruistic peers and a single seed with minimal upload capacity.  While earlier work has hinted at this result with
heuristics and simulations, it had proved difficult to come up with a provably stable scheme.  Although our original algorithm 
has drawbacks in terms of performance, we have suggested an improvement that allows trading locality for
performance.  Our proof was easily adapted to this case, which suggests that the methods presented here might allow for 
stability guarantees for other algorithms.

\section*{Acknowledgment}
The research of the authors was supported by the ARO
MURI grant W911NF-08-1-0233, ``Tools for the Analysis and
Design of Complex Multi-Scale Networks'', by the NSF
grant CNS-0910702, by the NSF Science \& Technology Center
grant CCF-0939370, ``Science of Information'' and by the Academy of Finland.



\bibliographystyle{IEEEtran}
\bibliography{IEEEabrv,k-chunk_stability}
%

\end{document}